\documentclass[12pt,a4paper]{article}
\pdfoutput=1
\usepackage[utf8]{inputenc}
\usepackage{epsf,amsmath,amssymb,graphicx,dcolumn}
\usepackage{caption}
\usepackage{subcaption}
\usepackage{scalefnt,ulem,pstricks}
\usepackage{booktabs,multirow,tabularx}
\usepackage{hyperref}
\usepackage{cleveref}
\usepackage{color}
\usepackage{rotating}
\usepackage{cancel}
\usepackage{microtype}
\usepackage[titletoc,title]{appendix}
\usepackage[numbers,sort&compress]{natbib}
\usepackage{amssymb}
\usepackage{bm, amsbsy}
\usepackage{xspace}

\providecommand{\href}[2]{#2}

\newcommand\as{\alpha_{\mathrm{S}}}

\def\to{\rightarrow}

\def\mw{m_W}

\newcommand{\pt}{\ensuremath{p_{\rm T}}}

\newcommand{\rcut}{\ensuremath{r_{\mathrm{cut}}}\xspace}
\newcommand{\ptveto}{\ensuremath{p_{T}^{\rm veto}}\xspace}

\newcommand{\TeV}{\ensuremath{\rm TeV}}

\newcommand\Matrix{{\sc Matrix}\xspace}
\newcommand\Munich{{\sc Munich}\xspace}
\newcommand\OpenLoops{{\sc OpenLoops}\xspace}
\newcommand\OpenLoopstwo{{\sc OpenLoops\,2}\xspace}
\newcommand\Recola{{\sc Recola}\xspace}

\newcommand{\fig}[1]{Figure~\ref{#1}}

\newcommand{\tab}[1]{Table~\ref{#1}}

\def\reffi#1{\mbox{Figure~\ref{#1}}}

\def\refta#1{\mbox{Table~\ref{#1}}}

\newcommand{\citere}[1]{\mbox{Ref.~\cite{#1}}}
\newcommand{\citeres}[1]{\mbox{Refs.~\cite{#1}}}

\renewcommand{\gg}{\ensuremath{gg}\xspace} 
\newcommand{\qqx}{\ensuremath{q\bar{q}}\xspace}
\newcommand{\qg}{\ensuremath{qg}\xspace}

\newcommand{\zz}{\ensuremath{ZZ}\xspace}
\newcommand{\ww}{\ensuremath{W^+W^-}\xspace}

\newcommand{\abbrev}{}

\newcommand{\LO}{\ensuremath{\rm{\abbrev LO}}\xspace}
\newcommand{\lo}{\LO}
\newcommand{\NLO}{\ensuremath{\rm{\abbrev NLO}}\xspace}
\newcommand{\nlo}{\NLO}
\newcommand{\NNLO}{\ensuremath{\rm{\abbrev NNLO}}\xspace}

\newcommand{\qqNNLO}{\ensuremath{q\bar{q}{\rm{\abbrev NNLO}}}\xspace}
\newcommand{\ggLO}{\ensuremath{gg{\rm{\abbrev LO}}}\xspace}
\newcommand{\ggNLO}{\ensuremath{gg{\rm{\abbrev NLO}}}\xspace}
\newcommand{\ggNLOgg}{\ensuremath{gg{\rm{\abbrev NLO}}_{gg}}\xspace}
\newcommand{\nNNLO}{\ensuremath{{\rm n{\abbrev NNLO}}}\xspace}
\newcommand{\nNNLOEW}{\ensuremath{{\rm n{\abbrev NNLO}_{EW}}}\xspace}

\newcommand{\nNNLOxEWqq}{\ensuremath{{\rm n{\abbrev NNLO}\!\times\!EW_{qq}}}\xspace}

\newcommand{\NNNLO}{\ensuremath{{\rm {\abbrev N^3LO}}}\xspace}

\newcommand{\Oas}[1]{\ensuremath{\mathcal{O}(\as^{#1})}\xspace}

\newcommand{\QCD}{{\abbrev QCD}\xspace}

\newcommand{\EW}{{\abbrev EW}\xspace}

\newcommand{\NLOQCD}{\NLO\!\QCD}
\newcommand{\NNLOQCD}{\NNLO\!\QCD}
\newcommand{\NLOEW}{\NLO\!\EW}

\newcommand{\elle}{\ensuremath{\ell}}

\newcommand{\mww}{\ensuremath{m_{WW}}}

\newcommand{\muR}{\ensuremath{\mu_{\rm R}}}
\newcommand{\muF}{\ensuremath{\mu_{\rm F}}}

\newcommand{\ptw}{\ensuremath{p_{{\rm T},W}}}
\newcommand{\ptwp}{\ensuremath{p_{{\rm T},W^+}}}
\newcommand{\ptwm}{\ensuremath{p_{{\rm T},W^-}}}

\newcommand{\ptll}{\ensuremath{p_{{\rm T},\ell\ell}}}

\newcommand{\ptlone}{\ensuremath{p_{{\rm T},\ell_1}}}

\newcommand{\ptmiss}{\ensuremath{p_{{\rm T,} {\rm miss}}}}
\newcommand{\dphill}{\ensuremath{\Delta\phi_{\ell\ell}}}

\newcommand{\njets}{\ensuremath{N_{\rm jets}}}

\begin{document} 
\hypersetup{pageanchor=false}
\begin{titlepage}
\renewcommand{\thefootnote}{\fnsymbol{footnote}}
\begin{flushright}
     MPP-2019-176\\
     ZU-TH 04/20
     \end{flushright}
\par \vspace{10mm}

\begin{center}
{\Large \bf $\boldsymbol{W^+W^-}$ production at the LHC: NLO QCD corrections\\[0.2cm] to the loop-induced gluon fusion channel}
\end{center}

\par \vspace{2mm}
\begin{center}
  {\bf Massimiliano Grazzini}$^{(a)}$, {\bf Stefan Kallweit}$^{(b)}$,\\[0.2cm]
  {\bf Marius Wiesemann}$^{(c)}$ and {\bf Jeong Yeon Yook}$^{(a)}$

$^{(a)}$ Physik-Institut, Universit\"at Z\"urich, 8057 Z\"urich, Switzerland 

$^{(b)}$ Dipartimento di Fisica, Universit\`a degli Studi di Milano-Bicocca and
INFN, Sezione di Milano-Bicocca, 20126, Milan, Italy

$^{(c)}$ Max-Planck Institut f\"ur Physik, F\"ohringer Ring 6, 80805 M\"unchen, Germany

\end{center}

\begin{center} {\bf Abstract} \end{center}\vspace{-1cm}
\begin{quote}
\pretolerance 10000

We compute the NLO QCD corrections to the loop-induced gluon fusion contribution in \ww production at the LHC.
We consider the full leptonic process $pp\to \ell^+\ell^{\prime\, -}\nu_{\ell}{\bar\nu}_{\ell^\prime}+X$,
by including resonant and non-resonant diagrams, spin correlations and off-shell effects.
Quark--gluon partonic channels are included for the first time in the calculation,
and our results are combined with NNLO predictions
to the quark annihilation channel at the fully differential level.
The computed corrections, which are formally of ${\cal O}(\as^3)$, 
increase the NNLO cross section by only about $2\%$, but have
an impact on the shapes of kinematical distributions, in part due to the jet veto,
which is usually applied to reduce the top-quark background.
Our results, supplemented with NLO EW effects, provide the most advanced fixed-order predictions available
to date for this process, and are compared with differential ATLAS data at $\sqrt{s}=13\,\TeV$.

\end{quote}

\vspace*{\fill}
\begin{flushleft}
February 2020

\end{flushleft}
\end{titlepage}
\hypersetup{pageanchor=true}

The production of vector-boson pairs plays a crucial role in precision studies 
at the Large Hadron Collider (LHC). 
Among the diboson processes, $W$-boson pair production has the largest 
cross section, and it provides direct access to the coupling between three
electroweak (\EW) vector bosons. This makes \ww production ideal to test
the gauge symmetry structure of \EW interactions and the mechanism of 
\EW symmetry breaking in the Standard Model (SM). It therefore provides 
an excellent probe of new-physics phenomena in indirect beyond-SM (BSM)
searches~\cite{ATLAS:2012mec,Chatrchyan:2013yaa,Wang:2014uea,Khachatryan:2015sga,Aad:2016wpd}%
\footnote{See \citeres{Frye:2015rba,Butter:2016cvz,Zhang:2016zsp,Green:2016trm,Baglio:2017bfe,Falkowski:2016cxu,Panico:2017frx,Franceschini:2017xkh,Liu:2019vid}
as examples of theoretical ideas to exploit precision in diboson processes to
constrain BSM physics.}, and it was instrumental for 
the Higgs boson discovery~\cite{Aad:2012tfa,Chatrchyan:2012xdj} as well as
measurements of its properties~\cite{Aad:2013xqa,ATLAS:2014aga,Aad:2015mxa,Aad:2016lvc,Aaboud:2018puo,Chatrchyan:2013mxa,Chatrchyan:2013iaa,Chatrchyan:2013lba,Khachatryan:2015mma,Khachatryan:2016vau,Khachatryan:2016vnn}.

The experimental precision of \ww measurements is continuously increasing. Its cross 
section has been measured first in proton--anti-proton collisions at the Tevatron~\cite{Aaltonen:2009aa,Abazov:2011cb},
and later in proton--proton collisions at the LHC with centre-of-mass energies
$\sqrt{s}=7$\,TeV~\cite{ATLAS:2012mec,Chatrchyan:2013yaa},
8\,TeV~\cite{Aad:2016wpd,Khachatryan:2015sga}, and
13\,TeV~\cite{Aaboud:2017qkn,CMS:2016vww,Aaboud:2019nkz}.
In particular, in the most recent 13\,TeV measurement of \citere{Aaboud:2019nkz} statistical uncertainties
have already reached the few-percent level, even for distributions in the fiducial phase space.
The systematic uncertainties are still comparably large, but they can be expected to further decrease in the future.

The increasing level of precision calls for continuous improvements 
in the theoretical description of \ww production at hadron colliders. The experimental
uncertainties can be matched on the theoretical side only by
highest-order computations in QCD and \EW perturbation theory.
At leading order~(LO) $W$-boson
pairs are produced through quark annihilation, which was calculated for 
on-shell $W$ bosons many years ago~\cite{Brown:1978mq}. 
Next-to-leading-order~(NLO) QCD predictions were first obtained in the on-shell 
approximation~\cite{Ohnemus:1991kk,Frixione:1993yp},
later incorporating leptonic $W$ decays with off-shell effects and spin
correlations~\cite{Campbell:1999ah,Dixon:1999di,Dixon:1998py,Campbell:2011bn}.
To reach the precision demanded by present \ww measurements, corrections
beyond \NLO QCD are indispensable.
NLO \EW corrections have been evaluated for on-shell $W$ bosons~\cite{Bierweiler:2012kw,Baglio:2013toa,Billoni:2013aba},
and with their full off-shell treatment~\cite{Biedermann:2016guo,Kallweit:2017khh,Kallweit:2019zez}. 
Although \EW effects have an impact of only few percent on the inclusive \ww rate, they 
can be significantly enhanced
up to several tens of percent at transverse momenta of about 1\,TeV.
At ${\cal O}(\as^2)$ the loop-induced gluon fusion channel provides a 
separately finite contribution, enhanced by the large gluon luminosity.
This contribution, which is part of the next-to-next-to-leading order (NNLO) corrections, has been known
for a long time~\cite{Glover:1988rg,Dicus:1987dj,Matsuura:1991pj,Zecher:1994kb,Binoth:2008pr,Campbell:2011bn,Kauer:2013qba,Cascioli:2013gfa,Campbell:2013una,Ellis:2014yca,Kauer:2015dma}.
The complete \NNLOQCD corrections were first evaluated in the on-shell approximation
for the inclusive cross section~\cite{Gehrmann:2014fva}, while the fully differential \NNLO predictions
for off-shell $W$ bosons were presented in \citere{Grazzini:2016ctr},
using the two-loop helicity amplitudes for $\qqx\to VV'$~\cite{Gehrmann:2014bfa,Caola:2014iua,Gehrmann:2015ora}.
Recently, the \NLO QCD corrections to the loop-induced gluon fusion contribution, which are formally of ${\cal O}(\as^3)$,
were evaluated~\cite{Caola:2015rqy} using the two-loop helicity amplitudes for 
$\gg\to VV'$ of \citeres{Caola:2015ila,vonManteuffel:2015msa},
considering only the gluon--gluon partonic channel.

One important aspect of the theoretical description of \ww production 
is the correct modelling of the jet veto
(see \citere{Jaiswal:2014yba,Meade:2014fca,Becher:2014aya,Monni:2014zra,Dawson:2016ysj} for example)
that is applied in \ww measurements to suppress the top-quark background.
Studies based on resummed computations at next-to-next-to-leading logarithmic (NNLL) accuracy
for the transverse momentum (\pt{}) of the \ww pair~\cite{Grazzini:2015wpa} and
for the jet-vetoed cross section~\cite{Dawson:2016ysj} matched to NNLO,
as well as the recent computation of \citere{Re:2018vac}
where NNLO predictions are matched to a parton shower, have shown that this is only relevant for small veto cuts.
Jet vetos of $30$\,GeV and higher used in recent \ww analyses~\cite{Aaboud:2019nkz} are 
sufficiently large to obtain reliable predictions for the fiducial cross section from fixed-order computations.
On the contrary, resummation effects are eventually
needed to obtain reliable predictions in the tails of some kinematical distributions,
for example in the invariant mass distribution of the \ww pair~\cite{Arpino:2019fmo}.

As pointed out in \citere{Grazzini:2018owa}, present experimental analyses for both \zz 
and \ww production assume the quark--antiquark~($\qqx{}$) annihilation and loop-induced 
gluon fusion (\gg{}) channels to be independent processes, although they mix in higher-order 
calculations through parton evolution.
Already at \NNLO there are diagrams that
mix the two production mechanisms, thereby suggesting that a unified treatment would be desirable.
Nevertheless, data are so far compared to {\it ad hoc} combinations of \NNLO calculations
for the quark annihilation channel and \NLO corrections to the loop-induced gluon fusion channel,
often by using $K$-factors (see e.g.\ \citeres{Aaboud:2017rwm,Aaboud:2017qkn,Sirunyan:2017zjc,CMS:2016vww}).
\citere{Grazzini:2018owa} made a decisive step for \zz production 
by combining the \NNLO calculation in the quark annihilation
channel with the \NLO calculation of the loop-induced gluon fusion channel.
This is particularly important to perform consistent variations 
of the renormalisation and factorisation scales, which in turn allows us
to obtain an estimate of perturbative uncertainties.

In this paper we take the same step for \ww production, and supplement predictions for
the \NNLOQCD cross section with \NLOQCD corrections to the loop-induced \gg{} 
contribution.  For the first time, we include also the (anti)quark--gluon (\qg{}) channel
entering the full \NLOQCD corrections to the loop-induced channel.
The combination of \NNLOQCD contributions with NLO corrections to the loop-induced gluon fusion channel provides
an estimate of the complete \NNNLO result and will be denoted by ``\nNNLO'' in the following.
This calculation constitutes
the most advanced perturbative \QCD prediction to date for the \ww process.
In order to reach the goal of percent-level theoretical uncertainties
for \ww production,  \QCD predictions must be combined with \EW corrections at least at NLO,
which has been achieved very recently in \citere{Kallweit:2019zez}.
Our calculation, including the aforementioned \EW corrections,
will be made publicly available in an updated version of \Matrix~\cite{Grazzini:2017mhc}.
In the following, we will discuss the details of the computation, analyse the effects 
of the newly computed corrections, and compare our best
phenomenological predictions for differential observables in the fiducial
phase space to ATLAS data at 13\,TeV.

We consider the process 
\begin{align}
\label{eq:process}
pp\to \ell^+\ell^{\prime\, -}\nu_{\ell}{\bar\nu}_{\ell^\prime}+X, 
\end{align}
where the charged final-state leptons and the corresponding (anti-)neutrinos have different flavours ($\ell\neq\ell^\prime$). 
All the resonant and non-resonant Feynman diagrams contributing
to this process are included, accounting for off-shell effects and spin correlations,
and our calculation is fully differential in the momenta of the final-state leptons and the associated QCD radiation.
The calculation is carried out by using the complex-mass scheme~\cite{Denner:2005fg}, without 
any resonance approximation.
Our implementation can deal with any combination of (massless) leptonic flavours,
$\elle,\elle^\prime\in \{e,\mu\}$, and in order to compare against
experimental data we focus on the process $pp\to \elle^+ \elle^{\prime -}\nu_\elle {\bar \nu}_{\elle^\prime}+X$
with $\elle,\elle^\prime=e$ or $\mu$ and $\elle\neq \elle^\prime$.
For the sake of brevity, we will denote this process as \ww production in the following.

\begin{figure}[t]
\begin{center}
\begin{tabular}{cccc}
\includegraphics[height=2.05cm]{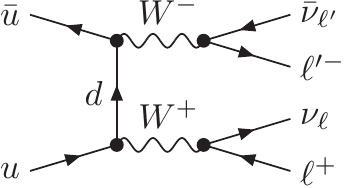} &
\includegraphics[height=2.05cm]{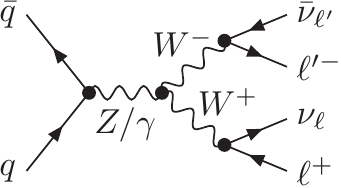} &
\includegraphics[height=2.05cm]{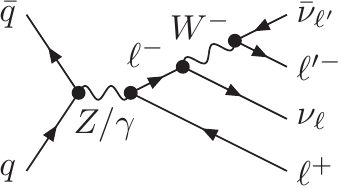} &
\includegraphics[height=2.05cm]{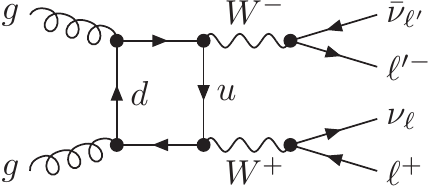} \\
(a) & (b) & (c) & (d) \\
\end{tabular}
\end{center}
\caption[]{\label{fig:diag}{Sample Feynman diagrams contributing to 
\ww production: (a-c) LO tree-level diagrams in the quark annihilation channel;
(d) loop-induced diagram of the gluon fusion channel entering at \Oas{2}.}}
\end{figure}

\begin{figure}[t]
\begin{center}
\begin{tabular}{c}
\includegraphics[height=2.5cm]{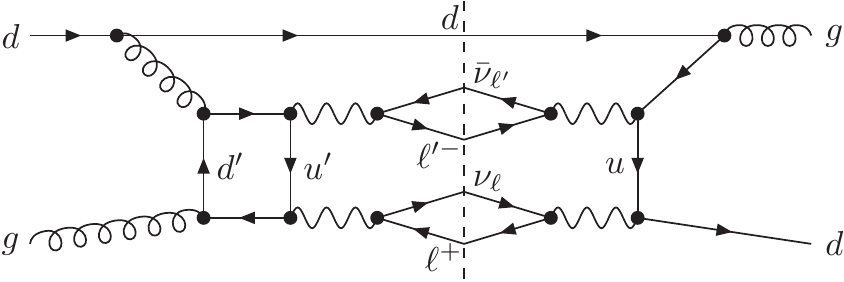} \\
\end{tabular}
\end{center}
\caption[]{\label{fig:mix}{Example of \NNLO interference between quark annihilation and loop-induced gluon fusion production mechanisms.}}
\end{figure}

Representative LO diagrams are shown in \fig{fig:diag}\,(a--c).
These are induced by \qqx{} annihilation and include 
$t$-channel \ww topologies~(panel a), 
$s$-channel \ww topologies~(panel b), and 
 $s$-channel Drell--Yan-type topologies~(panel c). 
\fig{fig:diag}\,(d), on the other hand, shows a \gg{} diagram
induced by a quark loop. The square of the corresponding amplitude enters at ${\cal O}(\as^2)$ and is part of the \NNLO corrections.
Furthermore, starting at \NNLO, there is a mixing between quark annihilation and gluon fusion contributions,
see \fig{fig:mix} for example. Such mixing renders a distinction between the two production mechanisms ambiguous.
  
Nevertheless, the square of the loop-induced \gg{} amplitude that enters 
at ${\cal O}(\as^2)$ is strongly enhanced by the large gluon luminosity,
and it is formally only \LO{} accurate. Consequently, the inclusion of the \NLO corrections to this contribution
becomes extremely important for obtaining a precise prediction for the \ww cross section.
Besides \gg{}-initiated diagrams at ${\cal O}(\as^3)$ the \NLO corrections entail
also \qg{}-initiated contributions.\footnote{There are 
also \qqx{}-initiated loop-induced contributions at $\mathcal{O}(\as^3)$. These 
are separately finite and completely negligible. Thus, we refrain from discussing them separately in the following, but
include them in our numerical predictions.}
The latter were neglected in the previous computation of the \NLO corrections~\cite{Caola:2015rqy}.
As we will see when presenting phenomenological results, the \qg{} contributions 
can become quite relevant, in particular in cases where a jet veto is applied.

\begin{figure}[t]
\begin{center}
\begin{tabular}{ccc}
\includegraphics[height=2.45cm]{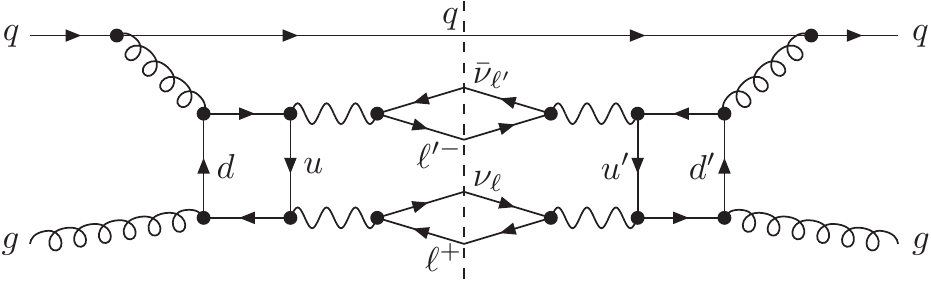}& \quad &
\includegraphics[height=2.45cm]{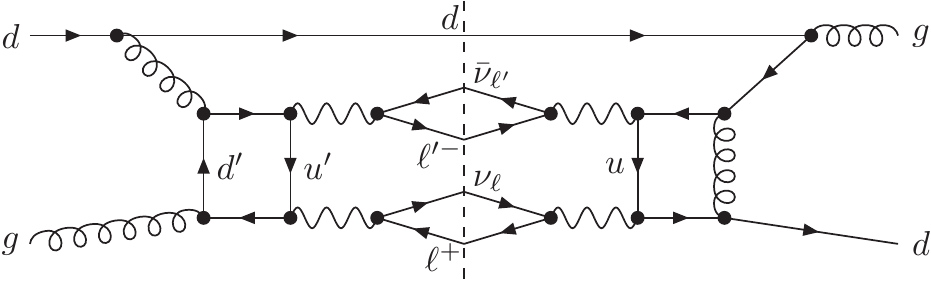} \\
(a) & \quad & (b) \\
\end{tabular}
\end{center}
\caption[]{\label{fig:qg}{Examples of \NNNLO contributions in the \qg{} channel, which are included~(a) and excluded~(b)
in our NLO calculation for the loop-induced gluon fusion channel.}}
\end{figure}

One should bear in mind that there are not only loop-induced contributions 
to the $gg$ channel at ${\cal O}(\as^3)$, and that we
only include contributions where both interfered amplitudes
contain a closed fermion loop, see \fig{fig:qg}\,(a) for an example and \fig{fig:qg}\,(b) for a counterexample.
At present it is not possible to consistently account for 
all the other contributions that would enter a complete next-to-next-to-next-to-leading-order~(\NNNLO) calculation.
For sufficiently inclusive observables, we expect the impact of the missing contributions at \NNNLO
to be within the perturbative uncertainties estimated from NNLO scale variations.
By contrast, NLO corrections to the loop-induced $gg$ channel are known to exceed
the uncertainties obtained through LO scale variations.
Our partial \NNNLO results (\nNNLO) 
therefore include all contributions up to $\mathcal{O}(\as^2)$,
together with the complete \NLO corrections to the loop-induced gluon fusion channel at $\mathcal{O}(\as^3)$.
As such, besides providing the maximum information available at present in \QCD
perturbation theory for this process,
our calculation provides a consistent estimate of the perturbative uncertainties
through renormalization and factorization scale variations.

We perform our calculation in the computational framework \Matrix~\cite{Grazzini:2017mhc}.
All tree-level and one-loop amplitudes are obtained with \OpenLoopstwo~\cite{Cascioli:2011va,Buccioni:2019sur}.%
\footnote{\OpenLoopstwo relies on its new on-the-fly tensor reduction~\cite{Buccioni:2017yxi} that
guarantees numerical stability in the entire phase space, especially in the IR-singular regions.
Within \OpenLoopstwo scalar integrals from {\sc Collier}~\cite{Denner:2016kdg} 
and {\sc OneLOop}~\cite{vanHameren:2010cp} are used.}
For validation of the loop-induced contribution, we have employed the independent matrix-element generator
\Recola~\cite{Actis:2016mpe,Denner:2017wsf}, finding complete agreement.
At two-loop level, we rely on the public C++ library \textsc{VVamp}~\cite{hepforge:VVamp} that implements 
the $\qqx\to VV'$ helicity amplitudes of \citere{Gehrmann:2015ora} and
the $\gg\to VV'$ helicity amplitudes of \citere{vonManteuffel:2015msa}.
The core of the \Matrix framework is the Monte Carlo program \Munich%
\footnote{\Munich is the abbreviation of ``MUlti-chaNnel Integrator at Swiss (CH) precision''---an
automated parton level NLO generator by S.~Kallweit. In preparation.}, which is capable of
computing both \NLO QCD and \NLOEW~\cite{Kallweit:2014xda,Kallweit:2015dum} corrections to arbitrary SM processes.
To reach \NNLOQCD accuracy, \Matrix uses
a fully general implementation of the $q_{\rm T}$-subtraction formalism~\cite{Catani:2007vq}.
The \NLO calculation is performed by using the 
Catani--Seymour dipole-subtraction method~\cite{Catani:1996jh,Catani:1996vz} and 
also with $q_{\rm T}$ subtraction~\cite{Catani:2007vq}, which provides an additional cross-check of our results.
\Matrix has been employed in a large number of NNLO calculations for colour-singlet processes at hadron
colliders~\cite{Grazzini:2013bna,Grazzini:2015nwa,Cascioli:2014yka,Grazzini:2015hta,Gehrmann:2014fva,Grazzini:2016ctr,Grazzini:2016swo,Grazzini:2017ckn,deFlorian:2016uhr,Grazzini:2018bsd}\footnote{It was
also used in the NNLL+NNLO computation of \citere{Grazzini:2015wpa}, and in the NNLOPS computation of \citere{Re:2018vac}.},
and it has been recently extended to heavy-quark production~\cite{Catani:2019iny,Catani:2019hip}.
\Matrix was also used for the NLO corrections to loop-induced \zz production~\cite{Grazzini:2018owa}.

We present predictions for $pp\to \elle^+ \elle^{\prime -}\nu_\elle {\bar \nu}_{\elle^\prime}+X$ production
with $\elle,\elle^\prime=e$ or $\mu$ and $\elle\neq \elle^\prime$ at the LHC with $\sqrt{s}=13$\,TeV.
For the \EW parameters we employ the $G_\mu$ scheme and set
$\alpha=\sqrt{2}\,G_F m_W^2\left(1-m_W^2/m_Z^2\right)/\pi$.
We compute the \EW mixing angle as $\cos\theta_W^2=(m_W^2-i\Gamma_W\,m_W)/(m_Z^2-i\Gamma_Z\,m_Z)$
and use the complex-mass scheme~\cite{Denner:2005fg} throughout.
The input parameters are set to the PDG~\cite{Patrignani:2016xqp} values: 
$G_F = 1.16639\times 10^{-5}$\,GeV$^{-2}$, $m_W=80.385$\,GeV,
$\Gamma_W=2.0854$\,GeV, $m_Z = 91.1876$\,GeV, $\Gamma_Z=2.4952$\,GeV,
$m_H = 125$\,GeV, and $\Gamma_H = 0.00407$\,GeV.
The on-shell masses of the bottom and the top quark are  $m_b = 4.92$\,GeV 
and $m_t = 173.2$\,GeV, respectively, and a top width of $\Gamma_t=1.44262$\,GeV is used.
The four-flavour scheme with $N_f=4$ massless quark flavours and massive bottom and
top quarks is used throughout. In this scheme contributions with massive bottom quarks in 
the final state are separately finite as they are regulated by the bottom-quark mass.
We follow the prescription adopted in \citeres{Gehrmann:2014fva,Grazzini:2016ctr} 
to avoid top-quark contamination: We drop all contributions with real
 bottom quarks in the final state to obtain a top-free \ww cross section.
Accordingly, we use the $N_f=4$ sets of the NNPDF31~\cite{Ball:2017nwa} parton 
distribution functions (PDFs) and choose the corresponding set for each perturbative order.\footnote{At LO there is no
$N_f=4$ NNPDF31 PDF set available, so we use the corresponding NNPDF30 set \cite{Ball:2014uwa}.}
The loop-induced gluon fusion contribution and also its \NLO corrections are 
always computed with \NNLO PDFs due to the lack of \NNNLO PDF sets.
To improve the perturbative treatment in the tails of the kinematical distributions,
we choose renormalization ($\muR$) and factorization ($\muF$) scales dynamically,
\begin{align}
\label{eq:dynscale}
\muR=\muF=\mu_0\equiv \frac12\,\left(\sqrt{m_{\elle\nu_\elle}^2+p^2_{{\rm T},\elle\nu_\elle}}
+\sqrt{m_{\elle^\prime\nu_{\elle^\prime}}^2+p^2_{{\rm T},\elle^\prime\nu_{\elle^\prime}}}\right),
\end{align}
where $m_{\ell\nu_\ell}$ and $p_{{\rm T},\ell\nu_\ell}$ ($m_{\elle^\prime\nu_{\elle^\prime}}$ and $p_{{\rm T},\elle^\prime\nu_{\elle^\prime}}$) are
the invariant masses and the transverse momenta of the reconstructed $W$ bosons.
Residual uncertainties are estimated from customary 7-point variations, i.e. 
by varying $\muR$ and $\muF$ around the central scale $\mu_0$ by a factor of two
while respecting the constraint $0.5\le \muR/\muF\le 2$.

Since the two-loop amplitudes including massive quarks are unknown to date, 
we employ the following approximation: The full dependence on the masses of top and bottom quarks is taken
into account everywhere in the calculation, except for the two-loop 
virtual contributions. Since two-loop $\qqx\to VV'$ diagrams with light fermion loops 
have a negligible impact, we expect also the massive quarks to induce subleading effects.
For the two-loop $\gg\to VV'$ contribution, on the other hand, we account for the heavy-quark masses 
approximately by reweighting the massless two-loop amplitude with the full \LO{} 
one-loop amplitude including the massive quarks.
Diagrams involving the Higgs boson are consistently included at each perturbative order, except for the two-loop contributions,
where we employ the same approximation as for the massive top-quark loops.

We follow the ATLAS analysis at $\sqrt{s}=13$\,TeV of \citere{Aaboud:2019nkz} and apply the selection cuts 
summarized in \refta{tab:cuts}.
The fiducial cuts involve standard requirements on the transverse momenta 
and pseudo-rapidities of the leptons, a lower invariant-mass cut and a lower cut
on the transverse momentum of the lepton pair, as well as a minimum requirement on the missing transverse momentum.
Most importantly, there is a jet veto as commonly applied in \ww measurements to 
suppress the top-quark background. However, a particular feature of the fiducial setup of \citere{Aaboud:2019nkz}
as compared to previous \ww analyses is the rather loose veto requirement which removes events with jets that have a $p_T$
larger than $\ptveto=35$\,GeV. Henceforth, our fixed-order calculation is
expected to be reliable except in phase-space regions that involve scales significantly larger than the jet veto scale.

\renewcommand{\baselinestretch}{1.5}
\begin{table}[t]
\begin{center}
\begin{tabular}{c}
\toprule
definition of the fiducial volume for $pp\to \elle^+ \elle^{\prime -}\nu_\elle {\bar \nu}_{\elle^\prime}+X$\\
\midrule
$p_{{\rm T},\elle/\elle^\prime}>27$\,GeV,\quad $|\eta_{\elle/\elle^\prime}|<2.5$,\\
$m_{\elle\elle^\prime}>55$\,GeV,\quad $p_{{\rm T},\elle\elle^\prime}>30$\,GeV,\quad $p_{{\rm T}, {\rm miss}}>20$\,GeV,\\
$\njets=0$\; for anti-$k_T$ jets \cite{Cacciari:2008gp} with $R=0.4$, $p_{T,j}>35$\,GeV, $|\eta_j|<4.5$\\
\bottomrule
\end{tabular}
\end{center}
\renewcommand{\baselinestretch}{1.0}
\caption{\label{tab:cuts}
Phase-space definitions of the \ww{} measurements by ATLAS at 13\,TeV~\cite{Aaboud:2019nkz}, with $\elle,\elle^\prime=e$ or $\mu$ and $\elle\neq \elle^\prime$.}
\end{table}
\renewcommand{\baselinestretch}{1.0}

We briefly introduce our notation for the various contributions:
\ggLO{} denotes the loop-induced gluon fusion channel contributing at ${\cal O}(\as^2)$,
while the \NNLO cross section in the quark annihilation channel (without the \ggLO{} contribution) is labelled \qqNNLO{}.
\ggNLO{} refers to the complete \NLO cross section of the loop-induced contribution, whereas
\ggNLOgg{} is its restriction to the \gg{}-induced partonic channel. Hence, the difference between these two predictions
corresponds to our newly computed contribution from the \qg{} channels. As discussed before, our partial \NNNLO result is
dubbed \nNNLO, and it includes the full \NNLO cross section supplemented by \NLO corrections 
to the loop-induced gluon fusion contribution at ${\cal O}(\as^3)$.

\renewcommand{\baselinestretch}{1.5}
\begin{table}[t]
\begin{minipage}{\textwidth}  
\begin{center}
\begin{tabular}{|l|rcr|rcr|}
\hline
\multicolumn{1}{|c|}{$\sqrt{s} = 13$ TeV} & jet veto  & & no jet veto & jet veto & & no jet veto \\
\hline
& \multicolumn{3}{c|}{$\sigma$\,[fb]} &\multicolumn{3}{c|}{$\sigma/\sigma_{\rm NLO}-1$} \\
\hline
\lo{}      
&      $    284.11(1)^{     +5.5 \%}_{     -6.5 \%}$	&	&      $    284.11(1)^{     +5.5 \%}_{     -6.5 \%}$	&                                     $    -15.5 \%$	&	&                                     $    -43.7 \%$	\\
\nlo{} 
&      $    336.42(3)^{     +1.6 \%}_{     -2.0 \%}$	&	&      $    504.36(3)^{     +4.1 \%}_{     -3.3 \%}$	&                                     $      +0. \%$	&	&                                     $      +0. \%$	\\
\qqNNLO{}
&      $     336.8(2)^{     +0.7 \%}_{     -0.5 \%}$	&	&      $     558.5(2)^{     +2.1 \%}_{     -1.9 \%}$	&                                     $     +0.1 \%$	&	&                                     $    +10.7 \%$	\\
\hline
& \multicolumn{3}{c|}{$\sigma$\,[fb]} &\multicolumn{3}{c|}{$\sigma/\sigma_{\rm {ggLO}}-1$} \\
\hline
\ggLO{}
&      $    21.965(4)^{    +25.7 \%}_{    -18.4 \%}$	&	&      $    21.965(4)^{    +25.7 \%}_{    -18.4 \%}$	&                                     $      +0. \%$	&	&                                     $      +0. \%$	\\
\ggNLOgg{}
&      $     31.68(6)^{    +10.8 \%}_{    -10.6 \%}$	&	&      $     38.49(6)^{    +15.9 \%}_{    -13.3 \%}$	&                                     $    +44.2 \%$	&	&                                     $    +75.2 \%$	\\
\ggNLO{}
&      $     28.79(6)^{     +7.8 \%}_{     -9.1 \%}$	&	&      $     37.57(6)^{    +15.3 \%}_{    -13.0 \%}$	&                                     $    +31.1 \%$	&	&                                     $    +71.0 \%$	\\
\hline
& \multicolumn{3}{c|}{$\sigma$\,[fb]} &\multicolumn{3}{c|}{$\sigma/\sigma_{\rm {NLO}}-1$} \\
\hline
\NNLO{}
&      $     358.7(2)^{     +1.2 \%}_{     -0.9 \%}$	&	&      $     580.5(2)^{     +2.9 \%}_{     -2.6 \%}$	&                                     $     +6.6 \%$	&	&                                     $    +15.1 \%$	\\
\nNNLO{}
&      $     365.6(2)^{     +0.4 \%}_{     -0.6 \%}$	&	&      $     596.1(2)^{     +2.8 \%}_{     -2.6 \%}$	&                                     $     +8.7 \%$	&	&                                     $    +18.2 \%$	\\
\hline
& \multicolumn{3}{c|}{$\sigma$\,[fb]} &\multicolumn{3}{c|}{$\sigma/\sigma_{\rm {nNNLO}}-1$} \\
\hline
\nNNLOEW{}
&      $     354.3(2)^{     +0.5 \%}_{     -0.8 \%}$	&	&      $     580.2(2)^{     +2.7 \%}_{     -2.6 \%}$	&                                     $     -3.1 \%$	&	&                                     $     -2.7 \%$	\\
\hline
\end{tabular}
\end{center}
\renewcommand{\baselinestretch}{1.0}
\caption{\label{tableincl} Fiducial cross sections at different perturbative orders and
relative impact on \NLO{} and \ggLO{} predictions, respectively. The quoted uncertainties correspond to scale variations as described in the text, and the numerical integration errors on the previous digit are stated in parentheses; for all ${\rm (n)NNLO}$ results, 
the latter include the uncertainty due the \rcut{} extrapolation~\cite{Grazzini:2017mhc}.}
\end{minipage}
\end{table}
\renewcommand{\baselinestretch}{1.0}

In \refta{tableincl} we present cross sections at the various perturbative orders corresponding to the fiducial phase space 
defined in \refta{tab:cuts}. Since the jet veto has a 
considerable impact on the structure of higher-order corrections, we show results in the same setup
both {\it with} and {\it without} jet veto (i.e., inclusive over QCD radiation).
The upper panel shows the QCD corrections to the quark annihilation channel,
the second panel focusses on the loop-induced gluon fusion contribution up to $\mathcal{O}(\as^3)$,
illustrating also the impact of the newly calculated \qg{} channels, and the third panel compares
the new best QCD prediction obtained in this paper, i.e.\ \nNNLO, with the \NNLO result.
Our best fixed-order prediction is labelled \nNNLOEW{} and is reported in the last panel of \refta{tableincl}.
It is obtained by combining our \nNNLO QCD result with NLO EW corrections
according to the prescription of \citere{Kallweit:2019zez}.%
\footnote{To be precise the \nNNLOEW{} prediction is obtained as follows.
Only the EW corrections that can be uniquely assigned to the \qqx production channel
are combined multiplicatively with the \qqNNLO{} result, whereas the remaining EW corrections
(photon-induced channels) as well as the \ggNLO{} contribution are added.
To account for the photon content of the proton, we use the
\texttt{NNPDF31\_nnlo\_as\_0118\_luxqed\_nf\_4} set~\cite{Bertone:2017bme} here,
and we apply the same EW input and renormalization schemes as in \citere{Kallweit:2019zez}.
By extending the notation of \citere{Kallweit:2019zez} this combined QCD--EW prediction could be also named \nNNLOxEWqq{}.}

The main conclusions that can be drawn from these results are the following:

\begin{itemize}
\itemsep0em
\item Cross sections with Born kinematics (LO and \ggLO{}) 
are obviously not affected by cuts on QCD radiation. By contrast, the jet veto has a significant impact 
on higher-order QCD corrections, both in the quark-initiated channels and for the loop-induced
gluon fusion contribution. In particular, the relative impact of the newly computed \qg{} channel 
becomes quite sizeable with the jet veto. This can be understood as follows: Inclusively, the \qg{} channel has a
rather small and negative impact, since the negative $\overline{\rm MS}$ mass factorization counterterm  balances 
the positive contribution from the real radiation matrix elements.
The jet veto acts only on the  real radiation. Thus, it removes a positive contributions from the \qg{} channel, rendering it more
negative and increasing its relative impact. As we will discuss below, also for kinematical distributions 
the jet veto has quite a strong impact on the \QCD corrections and the relative size of the \qg{} channel.
Without the jet veto radiative corrections are overall similar to those found for \zz production in \citere{Grazzini:2018owa}.
\item The \NLO (\NNLO) corrections relative to \LO (\NLO) in the quark annihilation channel amount to $+77.5\%$ ($+10.7\%$)
without jet veto, and they are reduced to $+18.4\%$ ($+0.1\%$) when the jet veto is applied.
\item \NLO corrections to the loop-induced gluon fusion channel are large: They are $+71.0\%$ in the case 
inclusive over QCD radiation and still $+31.1\%$ with the jet veto. The relative impact of the $\qg{}$ channel
can be extracted from the difference between the \ggNLO{} with \ggNLOgg{} predictions.
While in the inclusive case it reduces the \ggNLOgg{} corrections to the loop-induced gluon fusion channel by only roughly $6\%$,
this reduction increases to roughly $30\%$ when the jet veto is applied.
\item The \NNLOQCD effects with a jet-veto amount to $+6.6\%$ with respect to \NLO,
and they are almost entirely due to the \ggLO{} contribution,
whereas the latter accounts for only about one third of the $+15.1\%$ NNLO corrections without jet veto.
The impact of the \NLO corrections to the loop-induced contribution is to increase the \NNLO result by
about $1.9\%$~($2.7\%$) with~(without) the jet veto.
Excluding the \qg{} channels would increase the \nNNLO prediction with the jet veto by about $0.8\%$, 
while it has a subleading impact otherwise.
\item \NNLO and \nNNLO predictions without jet veto are fully compatible within scale uncertainties.
However, when the jet veto is applied, both at \NNLO and at \nNNLO the scale uncertainties are significantly reduced,
and the two predictions are marginally compatible. Such strong reduction of scale uncertainties can be seen as
a consequence of the improved convergence of the perturbative expansion,
but it also suggests that scale uncertainties should not be fully trusted
in this case as true perturbative uncertainties.
\item Contributions stemming from intermediate Higgs boson exchange are included,
but they are subdominant for the \ww selection cuts under consideration.
We find them to contribute only $-0.3$\% to the fiducial \ww cross section at \nNNLO. 
\item Comparing our predictions to the measurement of the fiducial cross section reported in
\citere{Aaboud:2019nkz}, $\sigma_{\rm fid}=379.1\pm 5.0 {\rm (stat)} \pm 25.4 {\rm (syst)} \pm 8.0 {\rm (stat)}$\,fb,
we note that \NNLO, \nNNLO and \nNNLOEW{} predictions are in agreement with the experimental result
within one standard deviation. In general, as far as the central value is concerned,
the positive effect of the \ggNLO{} corrections slightly improves the agreement,
while the EW corrections make the agreement slightly worse.
\end{itemize}

We now continue our presentation of phenomenological results by
discussing kinematical distributions. 
In the following, all figures are organized according to the same pattern:
The upper panel shows absolute cross sections at \LO{} (black, dotted), \NLO (red, dashed),
\NNLO (blue, dash-dotted) and \nNNLO (magenta, solid).
In the central panel the \NNLO and \nNNLO results including their scale uncertainty
bands are normalised to the central \NNLO prediction. 
The lower panel depicts the \NLO/\LO{} $K$-factors 
of the loop-induced gluon fusion contribution, with (\ggNLO{}; pink, solid) and without
(\ggNLOgg{}; brown, dash-double-dotted) the \qg{} contribution.
The results on the left are subject to the jet veto, whereas the ones on the right are inclusive over QCD radiation.

\begin{figure}[t]
\begin{center}                        
\begin{tabular}{cc}
\includegraphics[width=.43\textwidth]{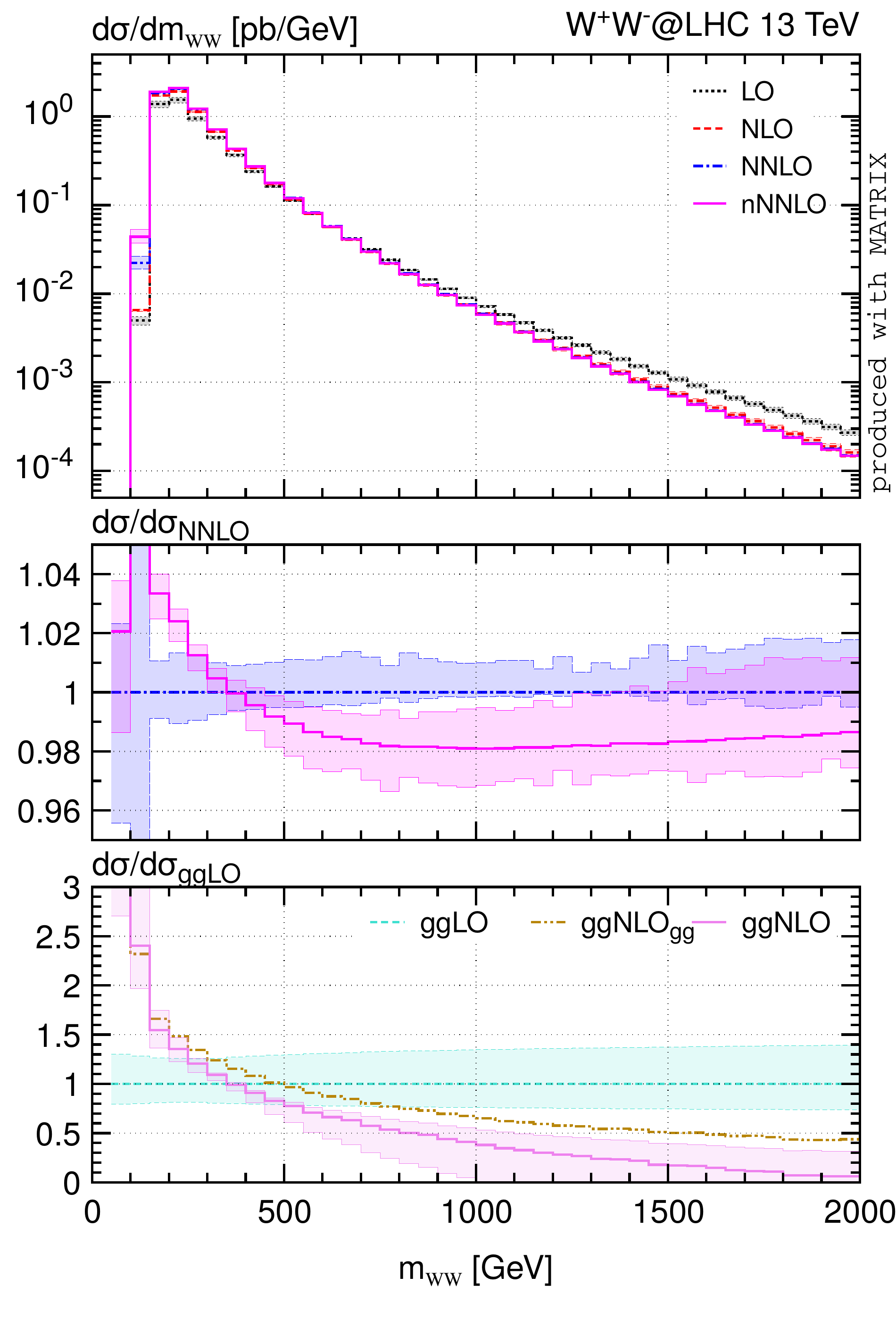} &
\includegraphics[width=.43\textwidth]{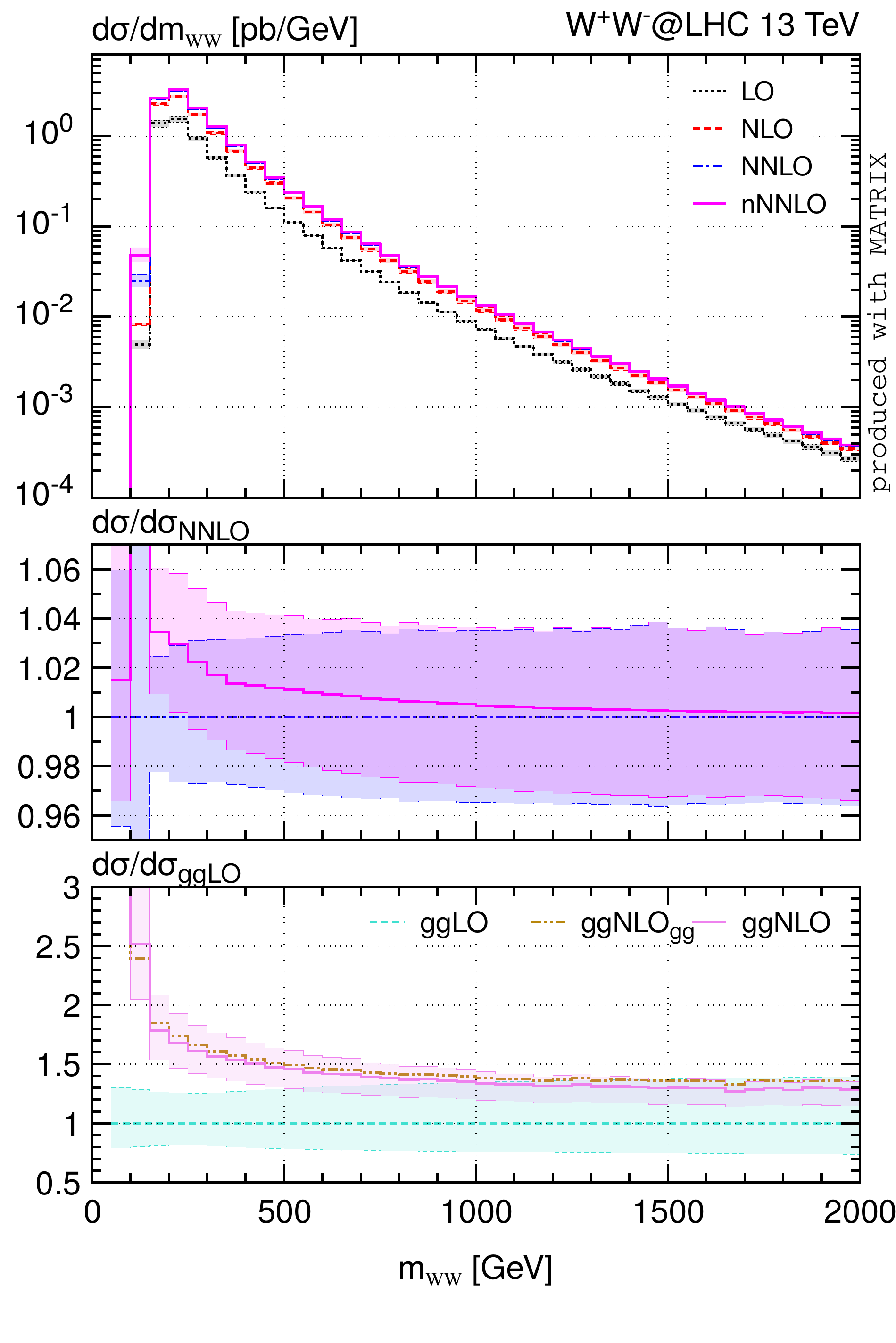}
\end{tabular}
\end{center}
\caption{\label{fig:mWW} Invariant-mass distribution of the \ww pair with jet veto (left) and without (right).}
\end{figure}      

We first show the invariant-mass distribution of the \ww pair (\mww{}) in \fig{fig:mWW}. 
It is interesting to notice how the relative corrections are affected by the jet veto.
Without jet veto (right panel) the loop-induced gluon fusion contribution at \NLO is largest 
at small invariant masses ($\sim +3\%$ at the peak with respect to \NNLO) and
continuously decreases towards larger \mww{} values, becoming negligible around 1\,TeV.
The \ggNLO{} correction is relatively flat (disregarding the region around the peak) and always positive,
while the \qg{} channel has a rather small impact, consistently with what we observed in \refta{tableincl}.
When the jet veto is applied (left panel), the shape of the \NNLO prediction is significantly affected.
The NLO corrections to the loop-induced contribution have a positive impact in the peak region ($\sim +3\%$ with respect to NNLO),
but at larger invariant masses their effect turns negative, reaching $\sim - 2\%$ in the TeV range.
The \ggNLO{} $K$-factor strongly decreases as \mww{} increases,
and the impact of the \qg{} channel is considerably large and negative at large \mww{}.

This behaviour is not unexpected. As it is well known, at large \mww{} the perturbative expansion is affected by large
logarithmic contributions of the form $\as^n \ln^{m} \mww{}/\ptveto$, which would require an all-order resummation.
Therefore, in the high-\mww{} region fixed-order predictions become unreliable, and
perturbative uncertainties are significantly larger than what can be inferred from customary scale variations.%
\footnote{Note that such pathological behaviour could be avoided by employing a 
dynamic jet-veto definition, as suggested for instance in \citeres{Kallweit:2017khh,Kallweit:2019zez}
to tame giant QCD $K$-factors in the high-energy tails of kinematical distributions.}
We note that the \nNNLO{} uncertainty band widens in the tail of the \mww{} distribution, becoming larger than the \NNLO{} band.
This effect is clearly driven by the interplay of the jet veto with the NLO corrections to the $gg$ channel.

\begin{figure}[t]
\begin{center}                        
\begin{tabular}{cc}
\includegraphics[width=.43\textwidth]{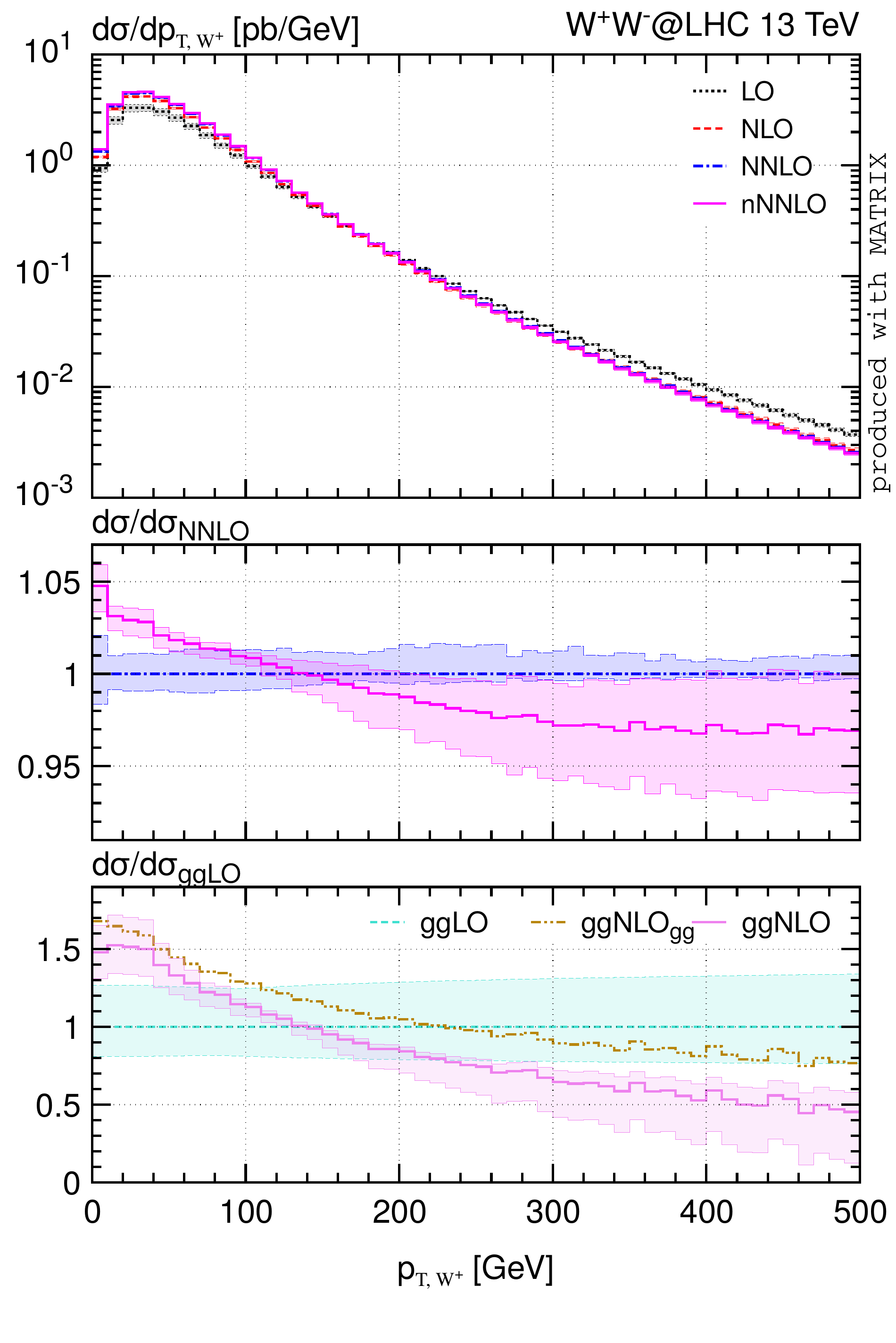} &
\includegraphics[width=.43\textwidth]{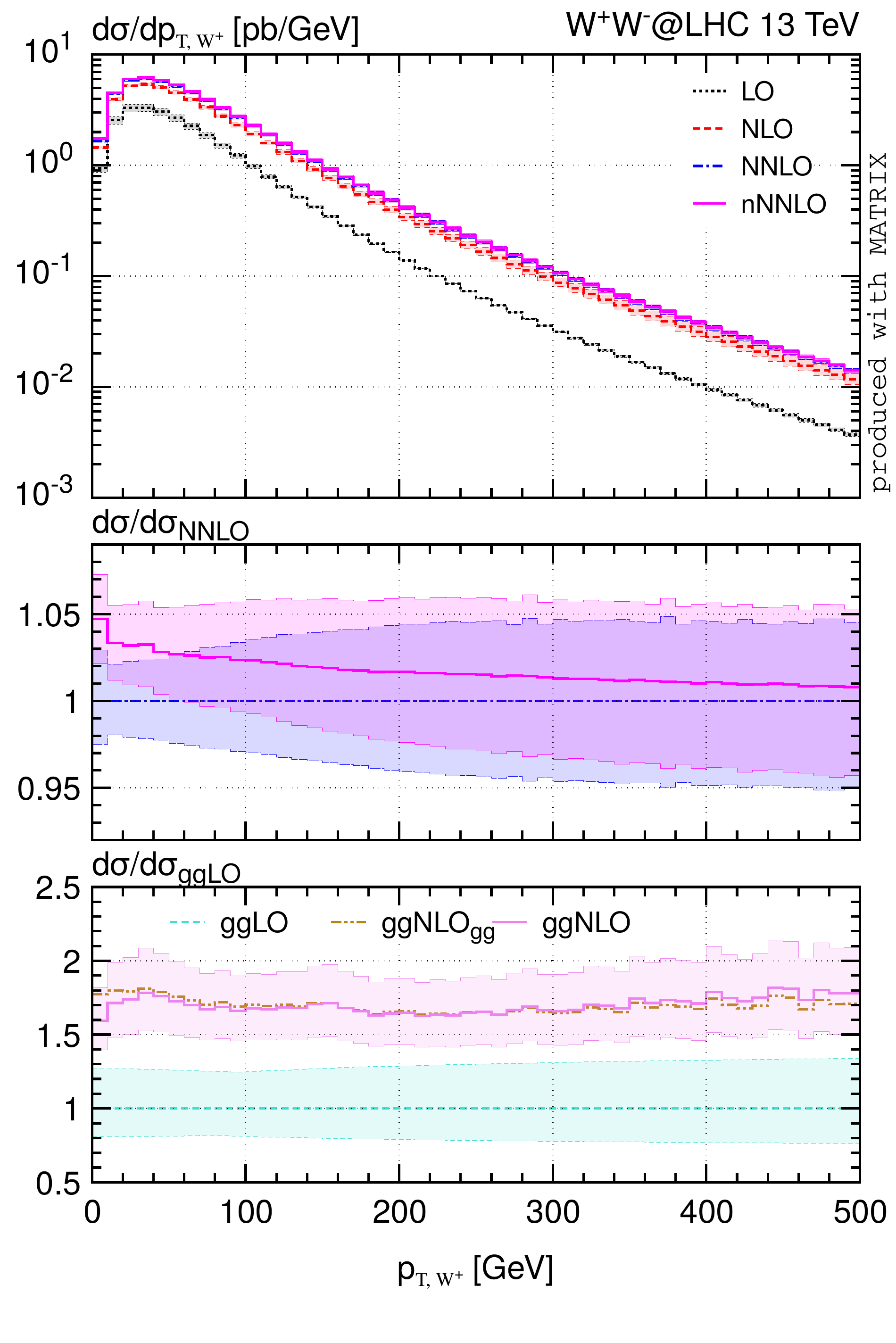}
\end{tabular}
\end{center}
\caption{\label{fig:ptW} Transverse-momentum spectrum of the $W$ boson with jet veto (left) and without (right).}
\end{figure} 

We find similar results for the transverse-momentum distribution of the reconstructed
  $W^+$ boson (\ptwp{}) in \fig{fig:ptW}; the distribution in \ptwm{} behaves similarly.
Without the jet veto (right panel) the \nNNLO corrections are positive at small transverse momenta,
and they decrease as \ptw{} increases, but remain positive and small in the tail of the distribution.
The \ggNLO{} $K$-factor is almost completely flat and around $+70\%$, with a minor impact of the \qg{} channel.
The small structure around $\ptw\sim 150\,$GeV is related to the massive-quark loop contributions.
When the jet veto is applied (left panel) the \nNNLO/\NNLO ratio is about $+5\%$ at small transverse momenta 
and steadily decreases until it reaches about $-3\%$ for $\ptw\gtrsim 300\,$GeV.
The \ggNLO{} correction is positive at small transverse momenta and it significantly decreases as \ptw{} increases,
with a considerable effect coming from the \qg{} channel.

\begin{figure}[t]
\begin{center}                        
\begin{tabular}{cc}
\includegraphics[width=.43\textwidth]{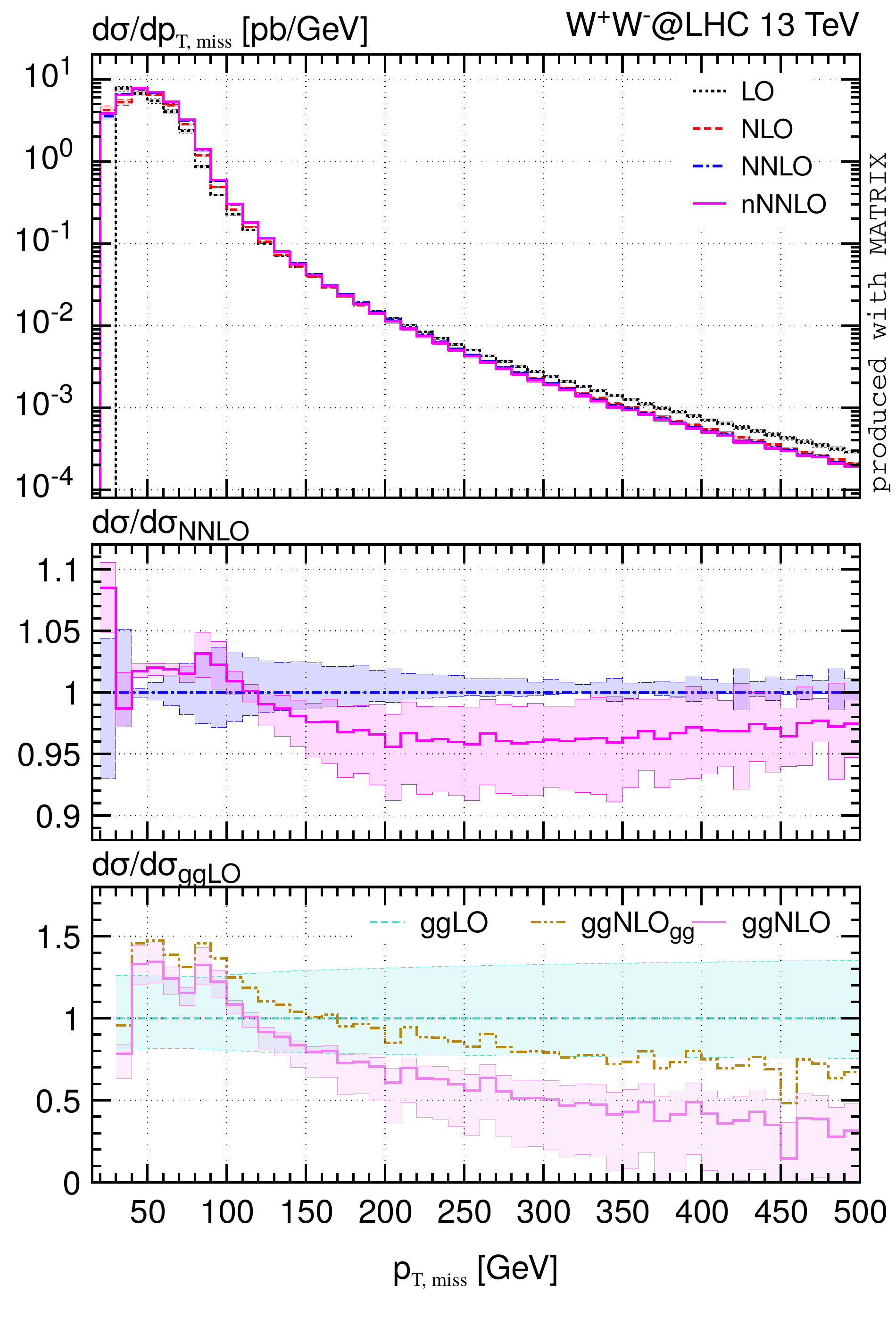} &
\includegraphics[width=.43\textwidth]{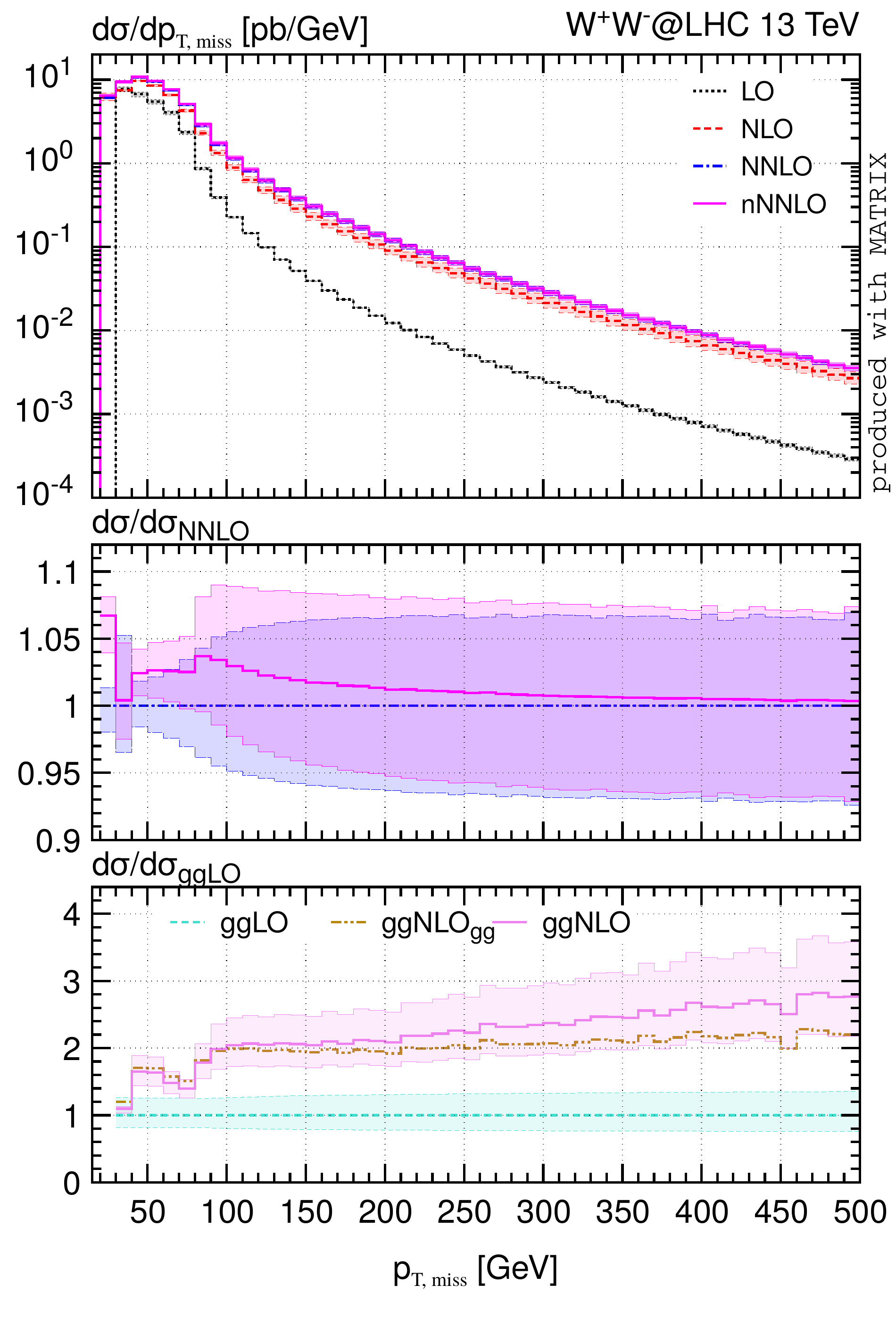}
\end{tabular}
\end{center}
\caption{\label{fig:pTmiss} Missing transverse-momentum distribution with jet veto (left) and without (right).}
\end{figure} 

In \fig{fig:pTmiss} we show the distribution in the missing transverse momentum ($\ptmiss$) computed from the vectorial 
sum of the neutrino momenta. The pattern of the corrections with and without jet veto is rather similar to the
two distributions considered before. We thus only discuss one peculiar additional feature,
namely the kink around $\ptmiss\sim \mw$, which is visible in all the predictions.
We also note that the size of the uncertainty bands for $\ptmiss>m_W$ clearly
increases. Such effect is due to the fact that at LO the two neutrinos are back-to-back with the leptons,
and therefore values of $\ptmiss$ larger than $m_W$ are strongly disfavoured as they
require at least one of the $W$ bosons to go off-shell. This can be seen also from 
the quickly decreasing LO result for $\ptmiss>m_W$ in the case in which no jet veto is applied.
At NLO the neutrinos can recoil against the additional jet, and this region receives a large correction.
Hence, the accuracy is effectively reduced by one order here, which explains also the enlarged uncertainty bands.
The impact of the NLO corrections in the quark annihilation channel
without jet veto is indeed huge, with $K$-factors of ${\cal O}(10)$ in the high-$\ptmiss$ region.

\begin{figure}
\begin{center}
\begin{tabular}{ccc}
\hspace*{-0.6em}\includegraphics[width=.32\textwidth]{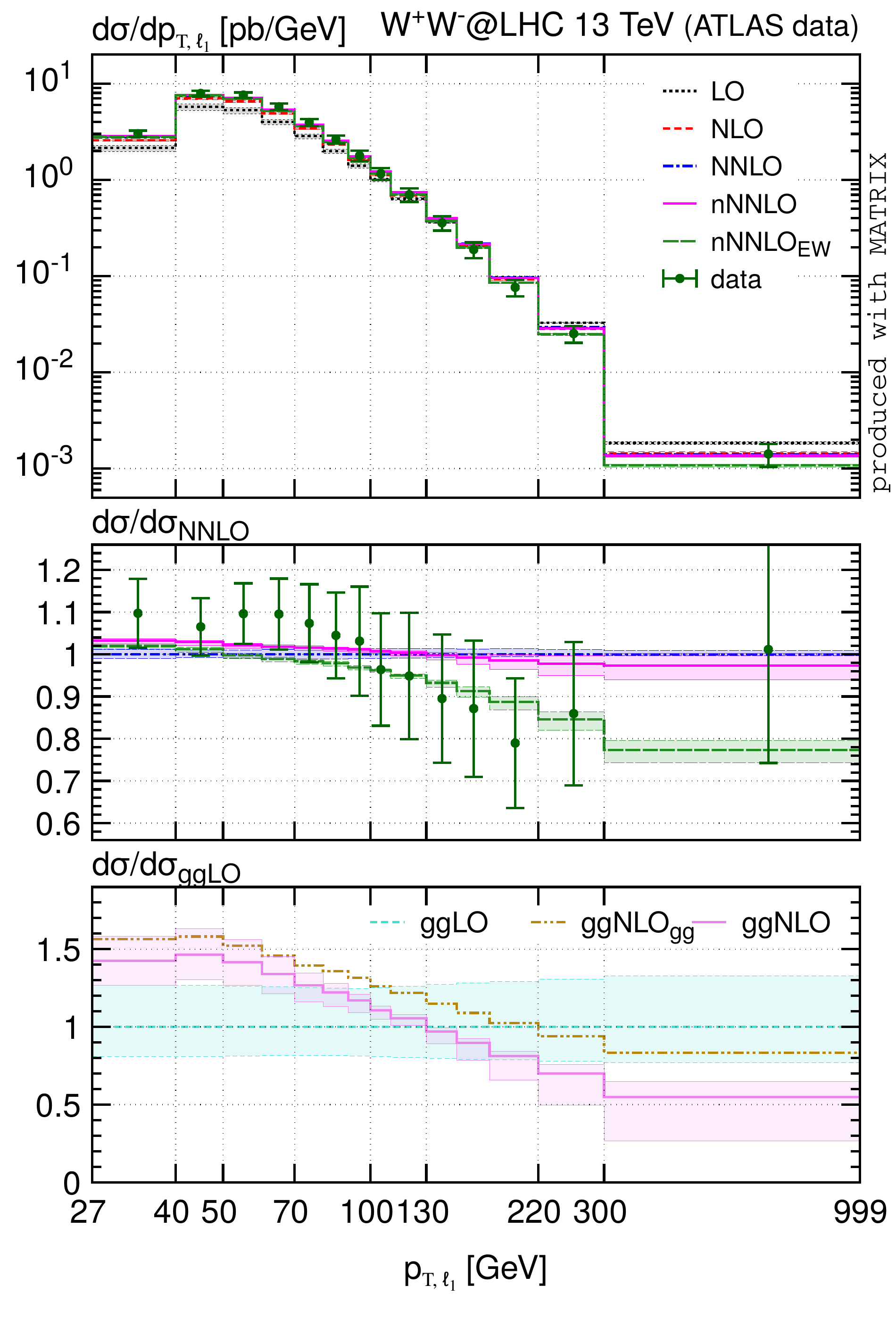} &
\hspace*{-0.6em}\includegraphics[width=.32\textwidth]{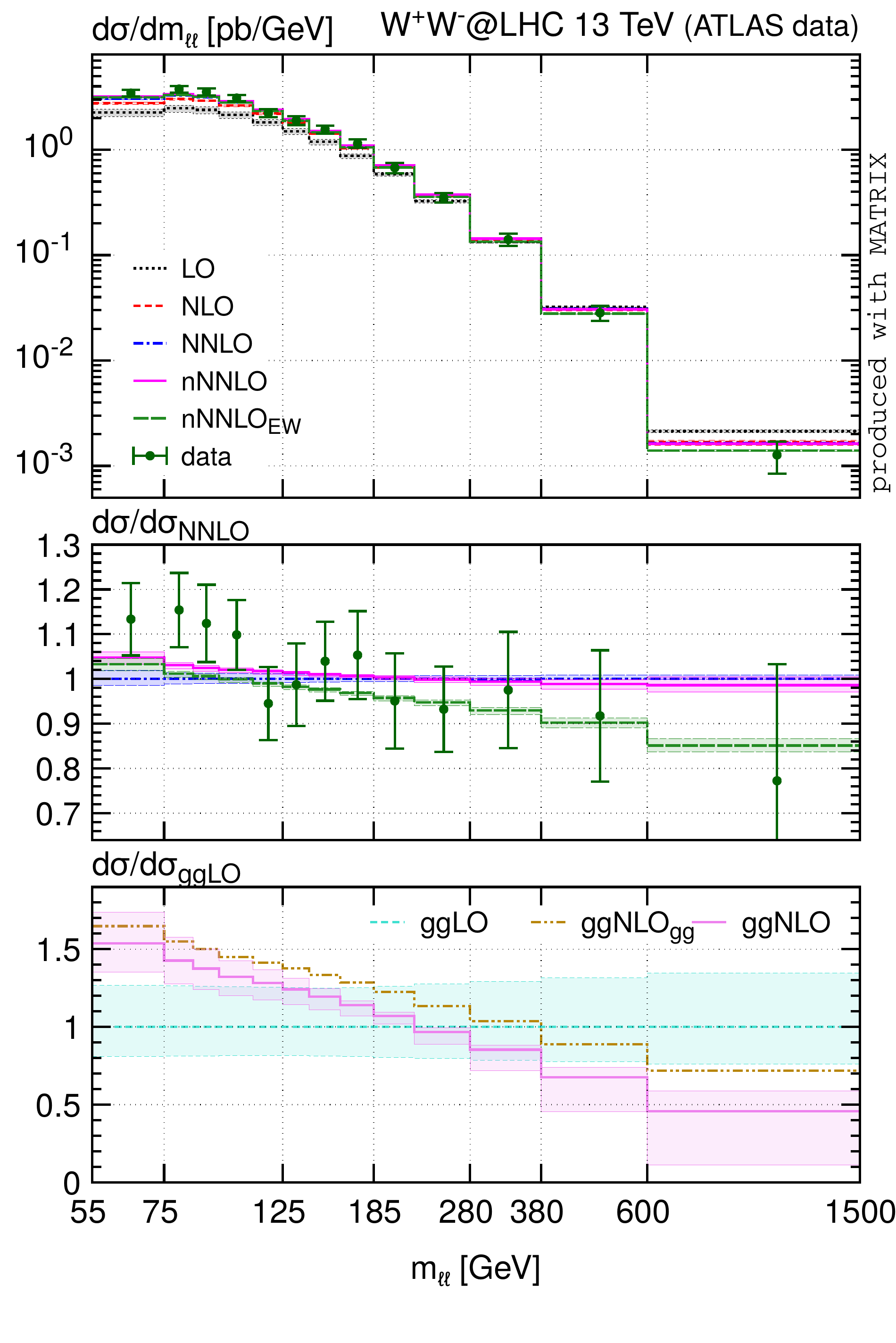} &
\hspace*{-0.6em}\includegraphics[width=.32\textwidth]{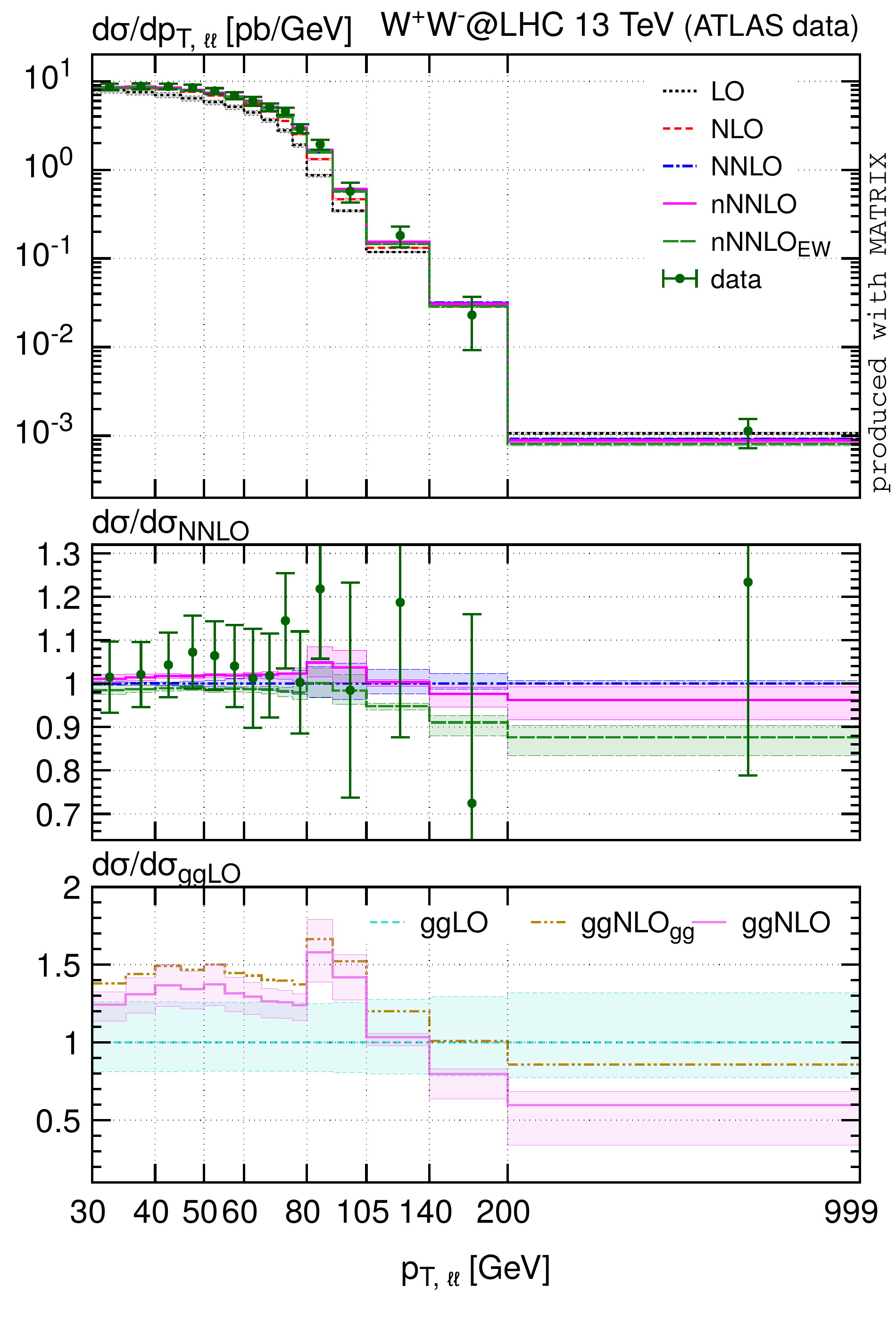}\\
\hspace*{-0.6em}\includegraphics[width=.32\textwidth]{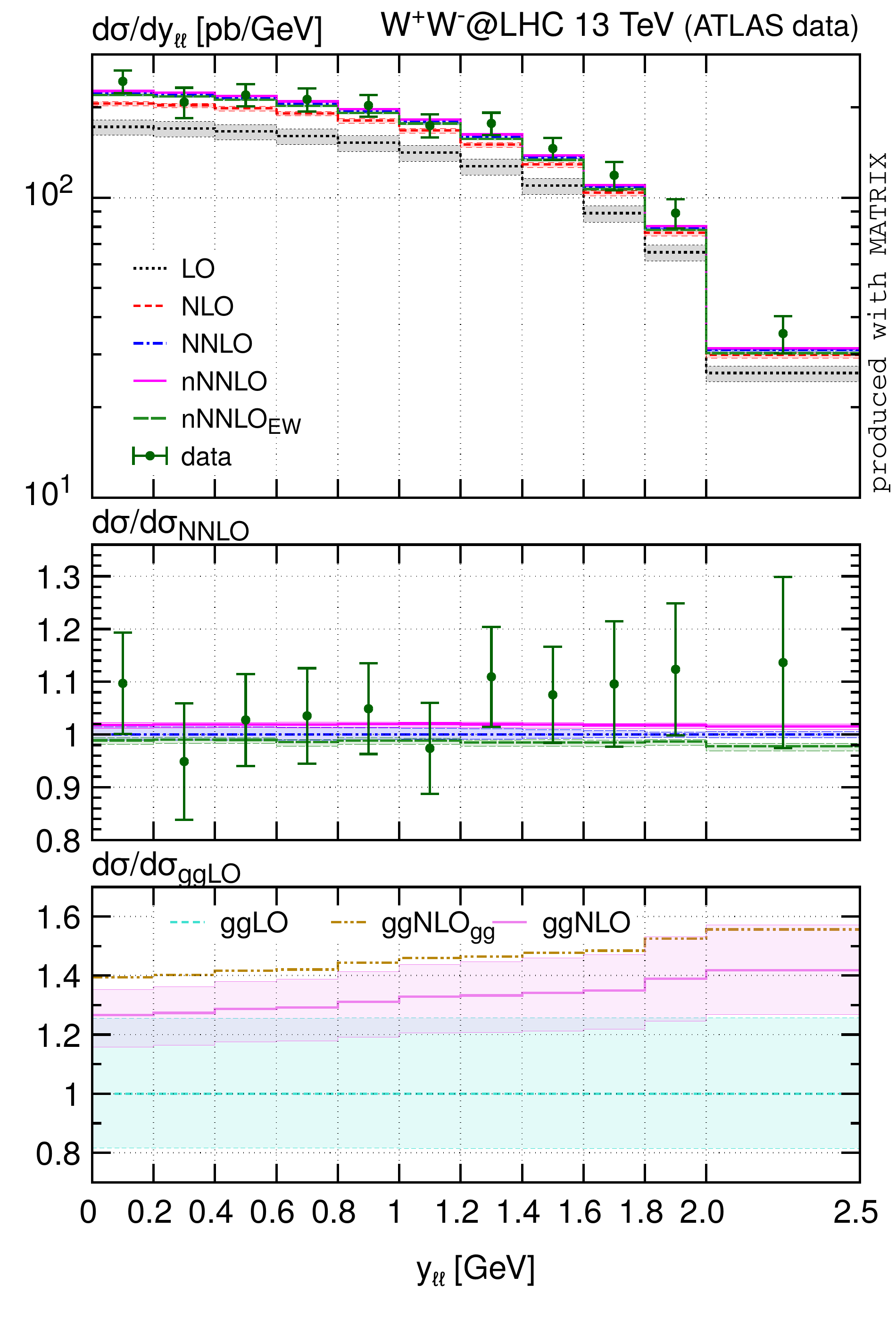} &
\hspace*{-0.6em}\includegraphics[width=.32\textwidth]{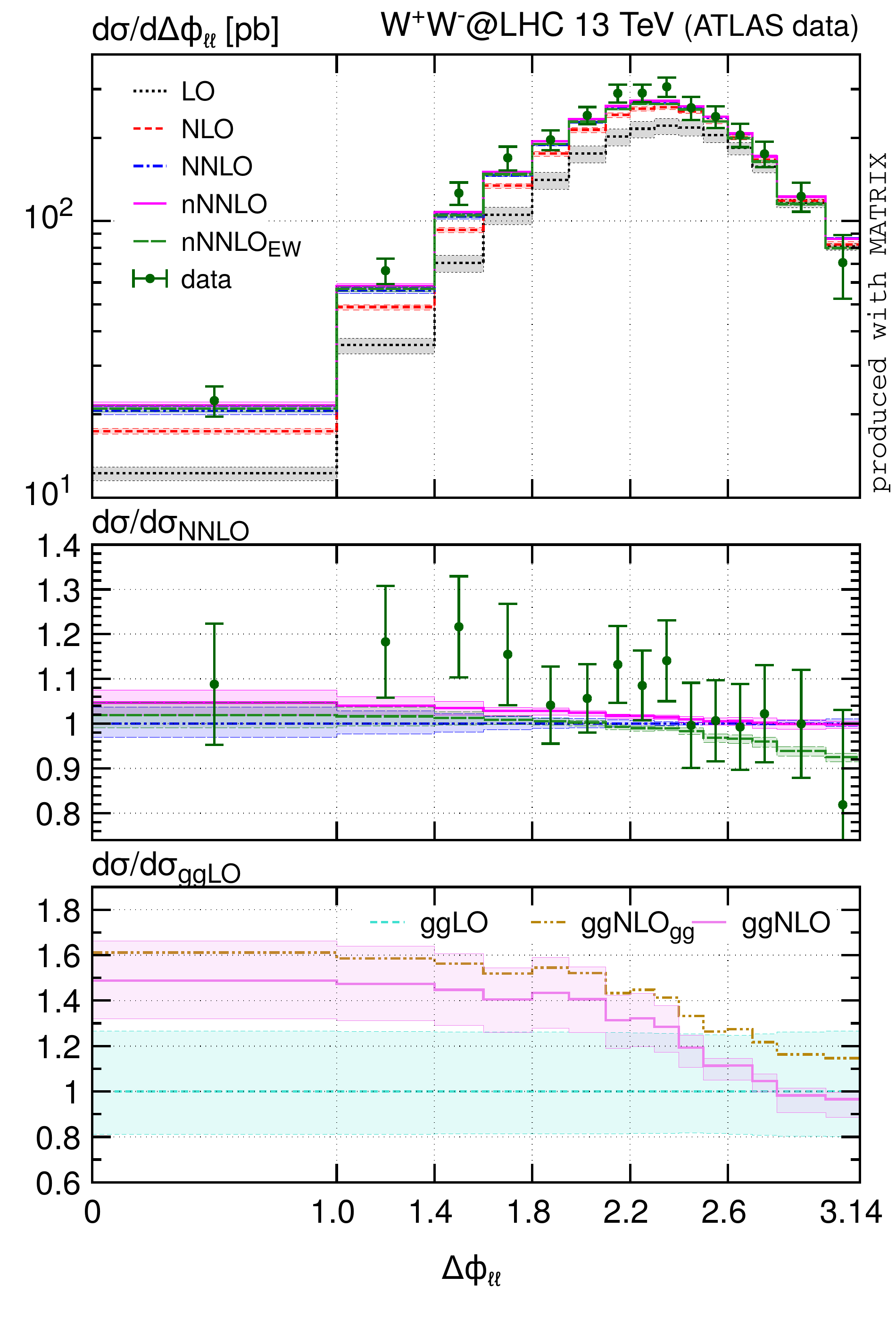} &
\hspace*{-0.6em}\includegraphics[width=.32\textwidth]{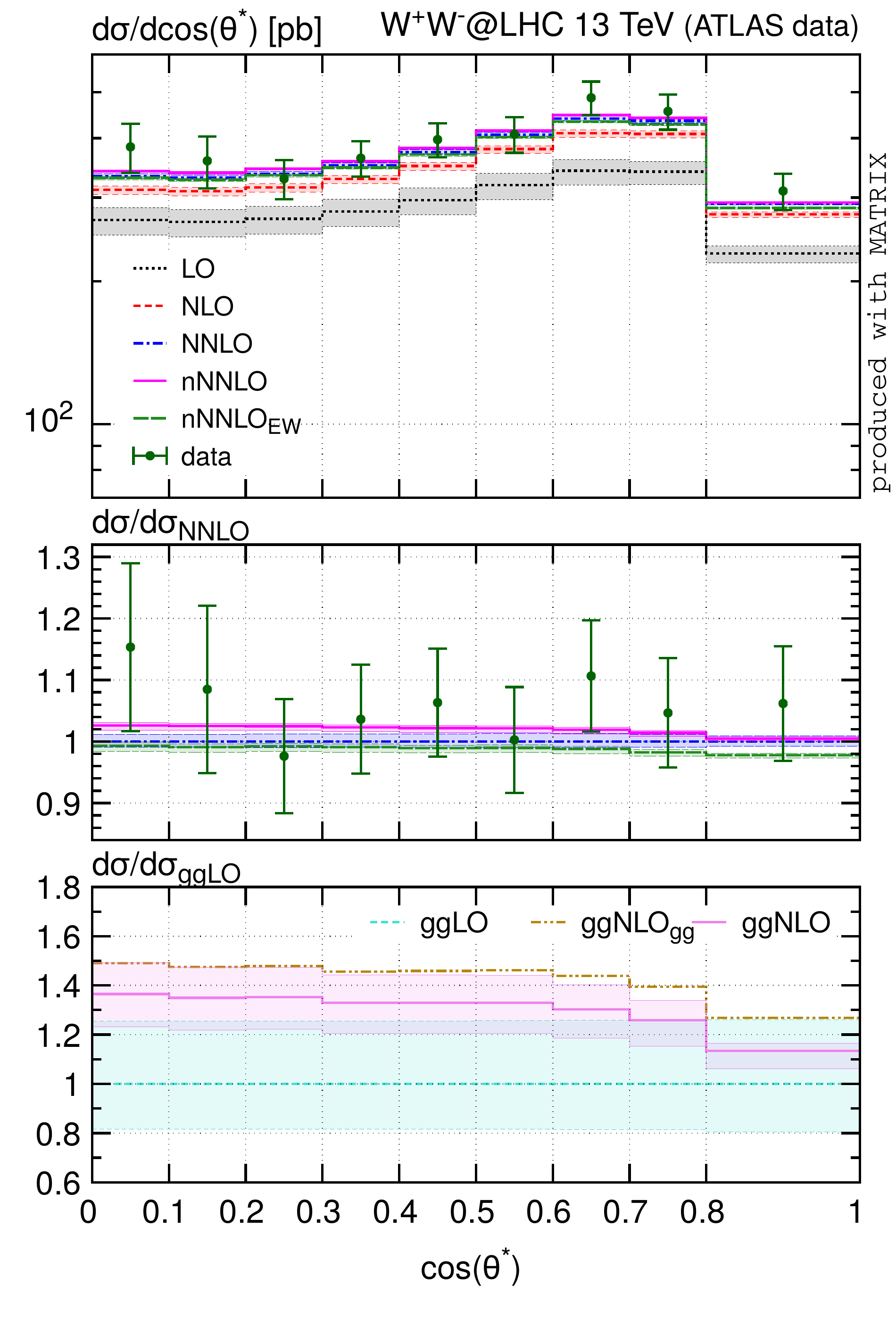}\\
\end{tabular}
\end{center}
\caption{\label{fig:theoryvsdata} Differential distributions in the fiducial phase space selections
of \tab{tab:cuts} compared to ATLAS 13\,TeV data~\cite{Aaboud:2019nkz};
top left: leading-lepton transverse-momentum distribution;
top center: lepton-pair invariant-mass distribution;
top right: lepton-pair transverse-momentum distribution;
bottom left: lepton-pair rapidity distribution;
bottom center: azimuthal distance between leptons;
bottom right: distribution in the variable $|\cos\theta^\ast|=|\tanh(\Delta\eta_{\ell\ell}/2)|$.}
\end{figure}  

We finally perform a comparison of our \nNNLO predictions to the ATLAS data of \citere{Aaboud:2019nkz}
in \reffi{fig:theoryvsdata}. On top of the pure QCD predictions, we also show the best available fixed-order result
\nNNLOEW{} (green, long-dashed), which was introduced in \refta{tableincl}
and includes also the EW effects calculated in \citere{Kallweit:2019zez}. 
In general, the NLO corrections to the loop-induced contribution slightly improve the agreement with the data,
especially when the \nNNLO prediction has a different shape compared to \NNLO, i.e.\ for the distributions
in the transverse momentum of the leading lepton (\ptlone{}, upper left plot),
the invariant mass of the lepton pair (\mww{}, upper central plot) and the
transverse momentum of the dilepton system (\ptll{}, upper right plot), as well as for the distribution
in the azimuthal angle between the leptons ($\dphill$, lower central plot).
The experimental uncertainties, however, are still too large to draw firm conclusions.
In the tails of the \ptlone{}, \mww{} and \ptll{} distributions, the negative effects of the EW Sudakov logarithms are
clearly more pronounced than those of the \ggNLO{} corrections.
For the rapidity distribution of the lepton pair ($y_{\ell\ell}$, lower left plot) and the distribution
in $|\cos\theta^\ast|=|\tanh(\Delta\eta_{\ell\ell}/2)|$ (lower right plot),
the \ggNLO{} corrections are largely flat with respect to the \NNLO result, and the better 
agreement of the \nNNLO distribution with data is only due to the normalization,
which was already found for the fiducial cross section in \tab{tableincl}.

We have calculated \NLOQCD radiative corrections to the loop-induced gluon fusion contribution in \ww production at the LHC.
Our predictions include, for the first time, also (anti)quark--gluon partonic channels.
By combining these results with the \NNLOQCD prediction for quark annihilation we have obtained 
an estimate of the full \NNNLO cross section, denoted by \nNNLO,
which represents the most advanced QCD prediction available to date for this process.
By considering the full leptonic signature $pp\to \ell^+\ell^{\prime\, -}\nu_{\ell}{\bar\nu}_{\ell^\prime}+X$,
we consistently account for off-shell effects, spin correlations and interferences.
We have presented phenomenological results for the fiducial cross section and distributions
applying the selection criteria used in the ATLAS measurement at $\sqrt{s}=13$\,TeV, which include a jet veto.
This jet veto, which is customarily applied to suppress the top quark background,
has a significant impact on the behaviour of radiative corrections.
In particular the shapes of important distributions are modified by \NLO corrections to the loop-induced gluon fusion
contribution. Our results are in good agreement with the experimental data, which, however, have still large uncertainties.
Our calculation has been implemented in the parton-level Monte Carlo program \Matrix
and will be made available in a forthcoming version of the code.
Together with the combination of EW corrections (already included in the comparison with experimental data here),
presented in parallel to this work in \citere{Kallweit:2019zez},
our results will boost the precision in physics studies based on the \ww signature at the LHC and future hadron colliders.

\noindent {\bf Acknowledgements.}
We thank Fabrizio Caola, Raul R\"ontsch and Lorenzo Tancredi for helpful discussions.
We are particularly indebted to Jonas Lindert and Federico Buccioni for useful correspondence, and for
for making private \OpenLoops amplitudes for the loop-induced channel available to us.
Moreover, we want to thank Jean-Nicolas Lang for his help on interfacing \Recola and validating the \OpenLoops results.
This work is supported in part by the Swiss National Science Foundation (SNF) under
contracts CRSII2-141847 and 200020-169041.
The work of MW was supported by the ERC Consolidator Grant 614577 HICCUP,
and that of SK by the ERC Starting Grant 714788 REINVENT.

\newpage
\renewcommand{\em}{}
\bibliographystyle{apsrev4-1}
\bibliography{wwnlogg}
\end{document}